\documentclass[11pt,twoside]{article} 
 
\usepackage{asp2004} 
\usepackage{epsf} 
\usepackage{graphicx}
\usepackage{lscape} 
 
\begin{document} 
\title{Six New ZZ Ceti Stars from the SPY and the HQS Surveys}   
\author{B. Voss$^1$, D. Koester$^1$, R. \O stensen$^2$, R. Napiwotzki$^3$,
  D. Homeier$^4$, and D. Reimers$^5$}   
\affil{$^1$Institut f\"ur Theoretische Physik und Astrophysik der
Universit\"at Kiel, Leibnizstra\ss e 15, 24098 Kiel, Germany}    
\affil{$^2$Instituut voor Sterrenkunde, K.U.Leuven, Celestijnenlaan 200D, 3001 Leuven, Belgium}
\affil{$^3$Centre for Astrophysics Research, University of Hertfordshire,
              College Lane, Hatfield AL10 9AB, UK}
\affil{$^4$Institut f\"ur Astrophysik, Georg-August-Universit\"at,
              Friedrich-Hund-Platz 1, 37077 G\"ottingen, Germany}
\affil{$^5$Hamburger Sternwarte,
              Gojenbergsweg 112, 21029 Hamburg, Germany}

\begin{abstract} 
We report on the discovery of six new ZZ Ceti stars. They were selected as
candidates based on preparatory photometric observations of
objects from the Hamburg Quasar Survey (HQS), and based on the spectra
of the Supernova Ia Progenitor Survey (SPY). Time-series photometry of 19
   candidate stars was carried out at the Nordic Optical Telescope (NOT) at
   Roque de Los Muchachos Observatory, Spain. The new variables are relatively bright, $15.4<B<16.6$. Among them is
   WD\,1150-153, which is the third ZZ Ceti star that shows photospheric CaII
   in its spectrum.
\end{abstract}

\section{Target Selection}
We selected candidate ZZ Ceti stars based on temperatures that we
derived from photometric data or spectra of the candidate objects, similar to the selection that is described
in \citet{voss}. 

About half of the targets are objects that are listed in the catalogue of
\citet{homeier} as probable white dwarfs in the HQS \citep{hagen}
survey. To derive accurate temperatures for these objects, we conducted photometric follow-up
observations of about 300 of these probable white dwarfs, using the four channel
CCD camera BUSCA at the
Calar Alto 2.2\,m telescope. The other half of the targets was selected based
on the high resolution spectra that were obtained at the ESO VLT with the UVES
spectrograph for the SPY project \citep{napiwotzki}.

We derived temperatures and gravities by fitting
the photometry or the spectra with synthetic photometry or synthetic spectra
that we determined from
  Koester WD model atmospheres. The resulting atmospheric parameters are
listed in Table \ref{table:1}. The typical uncertainties are about 0.07\,dex and 250\,K for the SPY data, or 0.25\,dex and 500\,K for the BUSCA-derived
parameters. Further details of the photometric observations and of the data
analysis can be found in \citet{voss}.

\begin{table}
\begin{minipage}[t]{\textwidth}
\caption{Properties of the observed ZZ Ceti candidate stars. Objects with parameters from
  BUSCA photometry are listed in the upper part of the table, those with
  parameters from SPY spectroscopy
  are given in the lower part. The last column gives the type of the objects,
  where NOV$x$ describes an object for which no variations with an amplitude
  higher than $x$\,mma were present. hDAV and cDAV refers to ZZ Ceti stars on the hot side and on the cool side of the instability strip, respectively.}             
\label{table:1}      
\begin{centering}                          
\renewcommand{\footnoterule}{}   
\smallskip
\begin{tabular}{llrcclr}        
\hline\hline                 
Object & \multicolumn{2}{c}{RA \hspace{2ex}  (J2000) \hspace{2ex}  DE}
& $B$& $T_{\mathrm{eff}}$& $\log g$ & type\\ 
 & & & (mag) & (K) & & \\
\hline                        
HS\,0210+3302 & 02 13 06.2 & +33 16 10 & 15.8 & 11924 & 7.39 & hDAV\\
HS\,0213+0359 & 02 15 36.7 & +04 13 38 & 16.6 & 13035 & 8.01 & NOV2\\
WD\,0235+069 & 02 38 33.1 & +07 08 10 & 16.6 & 10950 & 7.75 & cDAV\\
HS\,0733+4119 & 07 37 07.9 & +41 12 28 & 15.9 & 11162 & 7.72\footnotemark[1] & cDAV\\
HS\,1246+1232 & 12 49 02.3 & +12 16 14 & 15.5 & 12590 & 8.55 & NOV1\\
HS\,1701+5039 & 17 02 19.0 & +50 34 59 & 16.4 & 11586 & 8.05 & NOV2\\
HS\,1951+7147 & 19 50 45.6 & +71 55 39 & 16.8 & 11789 & 8.40 & NOV2\\
HS\,2217+2454 & 22 20 15.7 & +25 09 09 & 16.1 & 11731 & 7.74 & NOV3\\
HS\,2304+2809 & 23 06 36.1 & +28 25 30 & 16.3 & 12122 & 8.02 & NOV3\\
HS\,2351+3554 & 23 53 55.0 & +36 11 27 & 16.6 & 11325 & 7.58 & NOV2\\
\hline
WD\,0344+073 & 03 46 51.4 & +07 28 02 & 16.2 & 10470 & 7.77 & NOV2\\
HE\,0344$-$1207 & 03 47 06.7 & $-$11 58 09 & 15.8 & 11466 & 8.28 & cDAV\\
HS\,0401+1454 & 04 04 35.0 & +15 02 27 & 16.2 & 12375 & 8.10 & NOV2\\
WD\,0710+216 & 07 13 21.6 & +21 34 07 & 15.3 & 10222 & 7.97 & NOV1\\
WD\,1150$-$153 & 11 53 15.4 & $-$15 36 36 & 16.0 & 12453 & 8.03 & hDAV\\
HS\,1308+1646 & 13 11 06.1 & +16 31 03 & 15.5 & 10957 & 8.33 & NOV1\\
HS\,1641+1124 & 16 43 54.1 & +11 18 50 & 16.1 & 12209 & 7.96 & NOV1\\
WD\,1959+059 & 20 02 12.9 & +06 07 35 & 16.4 & 11033 & 8.23 & cDAV\\
WD\,2333$-$049 & 23 35 54.0 & $-$04 42 15 & 15.7 & 10506 & 8.00 & NOV2\\
\hline 
\end{tabular}
\end{centering}
\smallskip
$^1$\,\citet{homeier98} give $T_{\mathrm{eff}}=11420\,$K, $\log g=7.63$ for HS\,0733+4119.\\
\end{minipage}
\end{table}

\section{Time series observations and reduction}
Two runs of time series observations, with seven useful nights in total, were carried out at the 2.5\,m Nordic
Optical Telescope (NOT) at the Roque de los Muchachos Observatory, Spain, with
the instrument ALFOSC. This camera was used in a multi-window
fast-readout mode, with integration times between 30\,s and 50\,s, depending
on the target magnitude. The observations were made through a sky contrast filter (NOT \# 92) with a
bandwidth of 2750\,\AA, centered on 5500\,\AA. We observed all but one of the
variables for at least two hours, distributed over two adjacent nights.

\begin{figure}
\centering
\includegraphics[width=13cm]{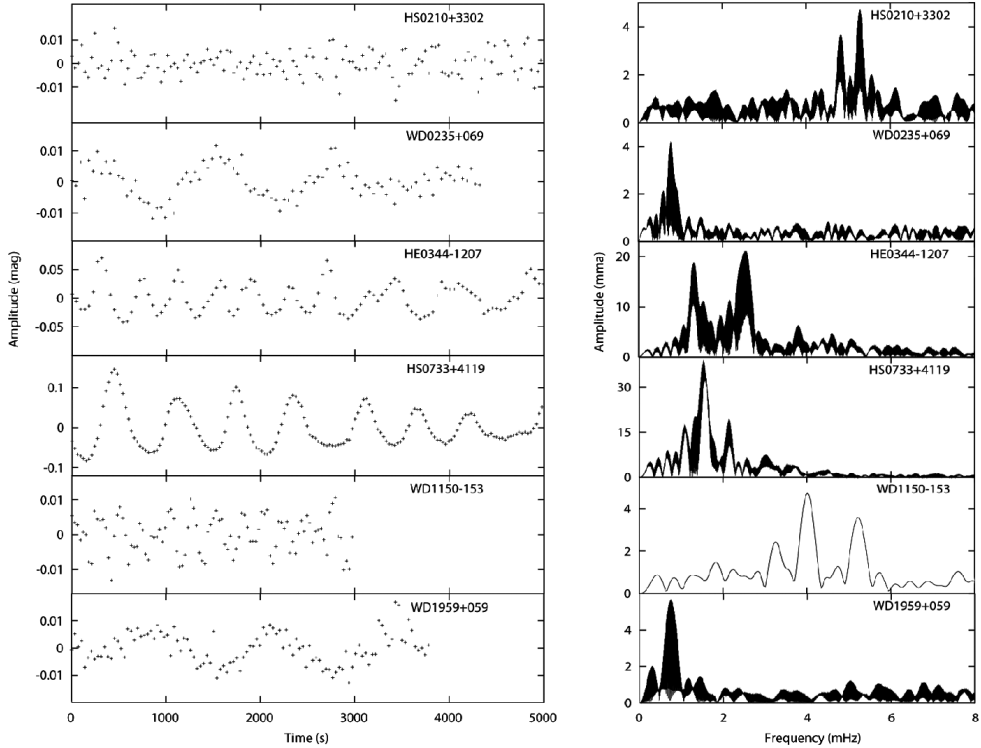}
\caption{The amplitude spectra and representative parts of the lightcurves of the new variables}
\label{fig1}
\end{figure}

Weighted differential aperture photometry was carried out using the program
RTP \citep{ostensen}, with an aperture size that was optimized for
the highest $S/N$. The resulting lightcurves and their amplitude spectra are shown in
Fig.\ref{fig1}; the pulsation frequencies and amplitudes are
listed in Table \ref{table:2}. The noise level in each FT was determined by
computing $\langle A \rangle$, the square root of the average power. It was determined from frequency regions of each FT which exclude the
ranges around the peak frequencies. 

\begin{table}
\begin{minipage}[t]{\columnwidth}
\renewcommand{\footnoterule}{}   
\begin{center}
\caption{Pulsational properties of the new variables}
\label{table:2}
\centering
\smallskip
\begin{tabular}{lcll}
\hline \hline
Object &  $\langle A \rangle$ (mma)& Main Amplitudes (mma)& Main Periods (s)\\
\hline
HS\,0210+3302 & 0.64 & 3.68, 4.74 & 207.5, 189.4\\ 
WD\,0235+069 & 0.38 & 4.21 & 1283.7\\
HE\,0344$-$0712 & 1.23 & 18.92, 11.37, 21.14 & 762.2, 461.0, 392.9\\
HS\,0733+4119 & 2.30 & 20.30, 38.73, 19.39 & 747.4, 656.2, 468.8\\
WD\,1150$-$153 & 0.58 & 4.73, 3.59 & 249.4, 191.7\\
WD\,1959+059 & 0.49 & 5.69 & 1350.4\\
\hline
\end{tabular} 
\end{center}
\end{minipage}
\end{table}

These six new variables, and the ten earlier ZZ Ceti detections that were made
based on our candidate selection work \citep{voss,cast,silvotti}, are about two magnitudes brighter than the majority of those
that are being discovered among the new white dwarfs detected by the SDSS.

\section{WD\,1150$-$153}
WD\,1150$-$153 has also been reported as a new ZZ Ceti star by
\citet{gianninas}, and is thus a parallel discovery. 

This star shows the CaII K line in its
spectrum. It is thus a DAZ WD \citep{koester05}, and only the third ZZ Ceti DAZ. This is of some interest as it has long been found that the shapes of the
H$\alpha$ line cores in the spectra of ZZ Ceti stars are too shallow to be
fitted by model spectra of non-rotating model atmospheres (e.g. \citet{koester98}). Thus, ZZ Ceti stars
seem to be relatively fast rotators, but the opposite is found from
asteroseismology, where the rotational splitting of pulsation modes provides
independent values of the rotation speeds that are close to zero
(e.g. Koester et al. 1998; Karl et al. 2005, and references therein). Horizontal surface
motions of the ZZ Ceti oscillations have been suggested as an alternative
explanation, and recent work by \citet{koester06} shows that such motions
can indeed quantitatively explain the observed line broadening. 

WD\,1150$-$153 exhibits a similar effect in the Ca II K line: \citet{berger} have
presented the first WD rotation velocities that are derived from the
broadening of the CaII line. WD\,1150$-$153 is one of very few objects for
which they determine a significant rotation velocity of $9.5 \pm
5.4$km\,s$^{-1}$. They suggest, however, that the line profile is probably not
caused by rotation but by the same effect that is responsible for the peculiar
H$\alpha$ line shapes in the ZZ Ceti stars.

\acknowledgements 
Based on observations made with the Nordic Optical Telescope, and partially
based on observations collected at the Centro Astron\'omico Hispano Alem\'an (CAHA) at Calar Alto, as well as on data obtained at the Paranal
       Observatory of the European Southern Observatory for programmes
       165.H-0588 and 167.D-0407. B.V. acknowledges support from the Deutsche
       Forschungsgemeinschaft under grant No. KO738/21-1, and by a travel grants from the OPTICON Trans-national
 access programme. 
 

\end{document}